\documentclass[10pt,conference]{IEEEtran}
\IEEEoverridecommandlockouts
\usepackage{cite}
\usepackage{amsmath,amssymb,amsfonts}
\usepackage{algorithmic}
\usepackage{graphicx}
\usepackage{textcomp}
\def\BibTeX{{\rm B\kern-.05em{\sc i\kern-.025em b}\kern-.08em
    T\kern-.1667em\lower.7ex\hbox{E}\kern-.125emX}}

\usepackage{xspace}
\usepackage{enumitem}
\usepackage{multirow}
\usepackage{colortbl}
\usepackage{array}
\ifCLASSOPTIONcompsoc
    \usepackage[caption=false, font=normalsize, labelfont=sf, textfont=sf]{subfig}
\else
\usepackage[caption=false, font=footnotesize]{subfig}
\fi
\usepackage{dblfloatfix}
\usepackage[most]{tcolorbox}
\usepackage{xcolor}
\usepackage{calc}
\usepackage{booktabs}
\usepackage{url}
\usepackage{booktabs}
\usepackage{makecell}
\newcommand{\tabitem}{~~\llap{\textbullet}~~}

\usepackage{color}
\definecolor{changeColor}{rgb}{0,0,255} 
\definecolor{deleteColor}{rgb}{255,0,0}
\definecolor{mygray}{RGB}{217, 217, 217}
\definecolor{myblue}{RGB}{191, 234, 234}
\definecolor{myred}{RGB}{255, 201, 196}
\definecolor{myred1}{RGB}{248, 113, 103}
\usepackage{marginnote}

\newcommand{\redtext}[2]{{\color{myred1}\marginnote{#1}#2}}

\AtBeginDocument{%
  \providecommand\BibTeX{{%
    \normalfont B\kern-0.5em{\scshape i\kern-0.25em b}\kern-0.8em\TeX}}}

\newcommand{\ie}{\textit{i}.\textit{e}.,\xspace}
\newcommand{\eg}{\textit{e}.\textit{g}.,\xspace}
\newcommand{\aka}{\textit{a.k.a.,}\xspace}

\newlength\MAX  

\begin{document}

\title{\emph{PairSmell}: A Novel Perspective Inspecting \\ Software Modular Structure
% \thanks{Identify applicable funding agency here. If none, delete this.}
}

\author{\IEEEauthorblockN{Chenxing~Zhong}
\IEEEauthorblockA{\textit{State Key Laboratory of Novel}\\ Software Technology\\
\textit{Nanjing University, China} \\
zhongcx@smail.nju.edu.cn}
\and
\IEEEauthorblockN{Daniel Feitosa}
\IEEEauthorblockA{\textit{Faculty of Science and Engineering} \\
\textit{University of Groningen} \\
the Netherlands\\
d.feitosa@rug.nl}
\and
\IEEEauthorblockN{Paris Avgeriou}
\IEEEauthorblockA{\textit{Faculty of Science and Engineering} \\
\textit{University of Groningen}\\
the Netherlands\\
p.avgeriou@rug.nl}
\and
\IEEEauthorblockN{Huang Huang}
\IEEEauthorblockA{
\textit{State Grid Nanjing}\\
Power Supply Company, China\\
sgcc.huang.huang@gmail.com}
\and
\IEEEauthorblockN{Yue Li}
\IEEEauthorblockA{\textit{State Key Laboratory of Novel}\\
Software Technology \\
\textit{Nanjing University, China}\\
yueli.dom@outlook.com}
\and
\IEEEauthorblockN{He Zhang}
\IEEEauthorblockA{\textit{ State Key Laboratory of Novel} \\
Software Technology \\
\textit{Nanjing University, China}\\
hezhang@nju.edu.cn}
}

\maketitle

\begin{abstract}

Enhancing the modular structure of existing systems has attracted substantial research interest, focusing on two main methods: (1) software modularization and (2) identifying design issues (\eg smells) as refactoring opportunities. However, re-modularization solutions often require extensive modifications to the original modules, and the design issues identified are generally too coarse to guide refactoring strategies. Combining the above two methods, this paper introduces a novel concept, \emph{PairSmell}, which exploits modularization to pinpoint design issues necessitating refactoring. We concentrate on a granular but fundamental aspect of modularity principles---\emph{modular relation (MR)}, \ie \emph{whether a pair of entities are separated or collocated}. The main assumption is that, if the actual MR of a pair violates its `apt MR', \ie an MR agreed on by multiple modularization tools (as raters), it can be deemed likely a flawed architectural decision that necessitates further examination.

To quantify and evaluate \emph{PairSmell}, we conduct an empirical study on 20 C/C++ and Java projects, using 4 established modularization tools to identify two forms of \emph{PairSmell}: inapt separated pairs $\mathit{InSep}$ and inapt collocated pairs $\mathit{InCol}$. Our study on 260,003 instances reveals that their architectural impacts are substantial: (1) on average, 14.60\% and 20.44\% of software entities are involved in $\mathit{InSep}$ and $\mathit{InCol}$ MRs respectively; (2) $\mathit{InSep}$ pairs are associated with 190\% more co-changes than properly separated pairs, while $\mathit{InCol}$ pairs are associated with 35\% fewer co-changes than properly collocated pairs, both indicating a successful identification of modular structures detrimental to software quality; and (3) both forms of \emph{PairSmell} persist across software evolution. This evidence strongly suggests that \emph{PairSmell} can provide meaningful insights for inspecting modular structure, with the identified issues being both granular and fundamental, making the enhancement of modular design more efficient.

\end{abstract}

% \begin{IEEEkeywords}
% pairwise modular smell, modularity, architectural smell
% \end{IEEEkeywords}

\section{Introduction}

\emph{Software Modularity} is an essential quality attribute reflecting how a system is structured into different parts (\emph{modules})~\cite{baldwin2000design}. 
This attribute has demonstrated a substantial impact on software reuse~\cite{kruger2020empirical}, and has been considered in various modern design scenarios, \eg microservices-based systems~\cite{abgaz2023decomposition} and LLM-enabled systems~\cite{wang2024reposvul}. 
Although the debate over \textit{``what constitutes a single module''} has sparked broad academic interests, determining appropriate modules is still challenging in practice. 
The reason is that modules can evolve quickly~\cite{schroder2021search}, due to updated business features and infrastructure technologies. Modules that worked well in the past might not fit into the current system.
Thus, a large part of the community's effort was spent on providing methodological support to improve the modularity of existing systems~\cite{teymourian2020fast,pourasghar2021graph,schroder2021search,candela2016using}, focusing primarily on two methods:

On one hand, \emph{software modularization} has been extensively investigated for nearly 30 years, with at least 143 papers published in the past decade~\cite{sarhan2020software}.
Relevant work~\cite{pourasghar2021graph,yang2022enhancing,mitchell2006automatic,teymourian2020fast}, treats modularization as an optimization problem, and searches for a (near-)optimal modular solution to replace the original modules.
Such solutions often ask for expensive changes to original systems, which might prevent developers from adopting them.
For example, even with refactoring effort as the optimization objective, a solution may introduce up to 170 move class~\cite{fowler2018refactoring} operations to a system~\cite{schroder2021search}.

On the other hand, some studies focus on identifying issues in modular structure, \eg based on quality metrics~\cite{garcia2021forecasting,bavota2013using,zhong2023measuring}, anti-patterns or smells~\cite{xiao2021detecting,mumtaz2021systematic,griffith2014design}. The issues are regarded as opportunities for refactoring in subsequent development, aiming at improving the degraded modules. The problem is that 
most of the issues are coarse at the module level, making it difficult for developers to determine refactoring strategies~\cite{CAI2023107322}.
A typical example is Cycle Dependency~\cite{fontana2017arcan}, where the chain of relations among several modules breaks the desirable acyclic nature of modules' dependency structure.
Although we know that cycle dependencies should be broken, it is difficult to decide which dependencies to break~\cite{oyetoyan2015decision}.

\begin{figure*}[t]
    \centering
    \subfloat[Dependency matrix of the system. ``X'' is a directional dependency between files.]{\includegraphics[width=.31\textwidth]{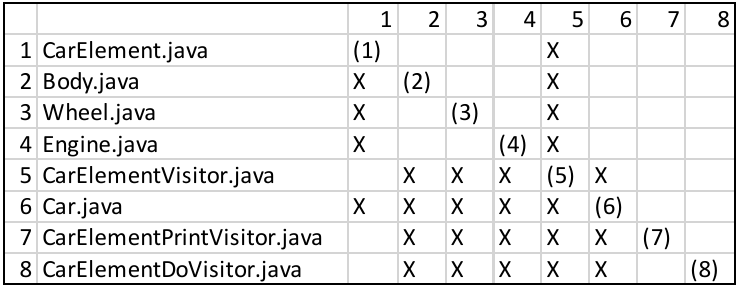}%\caption{fig1}
    }
    \quad
    \subfloat[Solutions from WCA (left) and FCA (right). ``O'' means two files collocated in the same module.]{\includegraphics[width=.18\textwidth]
    {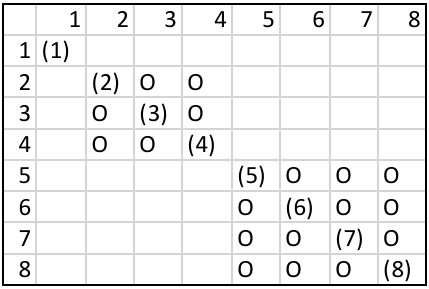}%\caption{fig1}
    \;
    \includegraphics[width=.18\textwidth]{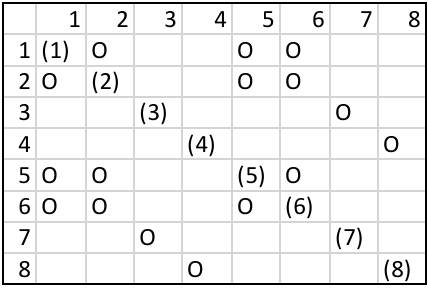}
    }
    \quad
    \subfloat[Consensus MRs.\colorbox{myblue}{Blue}: collocated;
    \colorbox{myred}{Red}: separated. ]{\includegraphics[width=.18\textwidth]{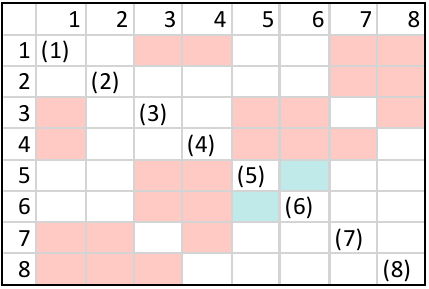}}%\caption{fig1}
    % \quad
    \caption{MRs agreed upon by multiple modularization tools are more reliable, as they comply with multiple rule sets.} 
    \label{fig: motivation case}
\end{figure*}

Our idea in this paper builds on both aforementioned methods. Rather than replacing the original modules, we propose that modularization tools can help identify issues that require refactoring. 
We focus on issues with a specific granularity: \emph{whether an entity (file) pair is collocated or separated within the same module}, termed \emph{\textbf{modular relation (MR)}}. This characteristic is central to several fundamental modularity principles like Common Closure Principle~\cite{martin2003agile} and Single Responsibility Principle~\cite{martin2018clean}. 
For instance, the Common Closure Principle suggests grouping entities that often change together~\cite{martin2003agile}. 
Moreover, entity pairs and their relationships are fundamental to many architecture analyses~\cite{almugrin2016using,czibula2019aggregated,benkoczi2018design}.
Our assumption is that \textit{if multiple modularization tools consensually design a MR as collocated or separated, it can be deemed a promisingly `apt MR'}, due to the consideration of diverse viewpoints.
This assumption, inspired by \emph{consensus clustering}~\cite{fred2005combining,zhang2022weighted} where similar cluster assignments indicate strong grouping between a pair of entities, reflects the consensus-based decision making in software development (\eg~\cite{tsoukalas2021machine,yang2023trust,muccini2018group}).
On the contrary, \textit{if the MR of a pair violates the apt one, this violation indicates an inappropriate architectural decision}~\cite{garcia2009identifying}, which we refer to as \textbf{\emph{Pairwise Modular Smell (PairSmell)}}.
In a nutshell, \emph{PairSmell} offers granular yet fundamental insights, helping developers inspect and improve software modules more effectively. It aims to identify issues necessitating refactoring based on multiple modularization tools, making development effort more targeted.

In this paper, we introduce, quantify, and evaluate \emph{PairSmell} as a novel type of issue for inspecting modular structure. 
To assess the severity of this issue, we conducted an empirical study involving 20 C/C++ and Java projects from GitHub. 
To support this study, we developed a tool, integrating 4 established modularization tools, to automatically detect \emph{PairSmell} from the modular structure of a development architectural view.
We mined 22,528
code commits across 473 diverse snapshots, and inspected 146,668,710 separated and 3,866,940 collocated pairs of entities. Based on the dataset, our study identified 260,003 \emph{PairSmells}, including 73,536 inapt separated pairs (\aka $\mathit{InSep}$) and 186,467 inapt collocated pairs ($\mathit{InCol}$).

The empirical results reveal that: (1) \emph{PairSmell} is prevalent among projects, with $\mathit{InSep}$ and $\mathit{InCol}$ instances covering 14.60\% and 20.44\% entities on average; (2) on average, entities in $\mathit{InSep}$ MRs co-change 190\% more than in other  separated pairs, and entities in $\mathit{InCol}$ MRs co-change 35\% less than other collocated pairs, dramatically deviating from well-structured modules; (3) \emph{PairSmell} persists in software projects if left unaddressed, where the percentages of $\mathit{InSep}$ and $\mathit{InCol}$ pairs remain stable as systems grow. 
In summary, our study makes the following contributions:

\begin{enumerate}
    \item A novel type of architectural smell and its identification approach are proposed, enabling the revelation of granular yet fundamental modular issues.
    
    \item An empirical study on the architectural smell is present, revealing that such smells are prevalent but detrimental to software maintenance and change, and could persist for long if left unaddressed.

    \item The novel smell is discussed, and its implications for practice and future research are illustrated.

    \item The benchmarks and replication package collected from 20 open source projects are publicly available~\cite{replication} for continued research of the novel type of smell.
\end{enumerate}

\section{Pairwise Modular Smell}
\label{sec: Pairwise Modular Smell}

Before proposing \emph{PairSmell}, we first illustrate its assumption inspired by \emph{consensus clustering}~\cite{fred2005combining,zhang2022weighted}, where similar cluster assignments for a pair of entities, indicate that these entities should be grouped together.
Specifically, a simple Java system from~\cite{jin2020exploring} is modularized using two widely used modularization tools, WCA~\cite{maqbool2004weighted} and FCA~\cite{teymourian2020fast} (detailed in Section~\ref{sec: tool implementation}), as shown in Fig.~\ref{fig: motivation case} (a) and (b).
Comparing these resulting solutions, we found promising insights in cases where both solutions consensually design the MR of a pair, as Fig.~\ref{fig: motivation case} (c).
An instance of collocated MR is found between \emph{CarElementVisitor} (row 5) and \emph{Car} (6).
By inspecting the architecture in Fig.~\ref{fig: motivation case} (a), we notice that these two files are structurally connected, featuring two direct dependencies and multiple indirect dependencies (\ie via files in row 2, 3, and 4).
Another example is a separated MR between \emph{CarElement} (1) and \emph{Wheel} (3), where we observe only a direct dependency. This separated MR appears justifiable, given the densely connected nature of this architecture.

To summarize, this example illustrates the rationale behind using MRs agreed by multiple modularization tools as promisingly apt MRs. It is crucial to recognize that these MRs are not infallible; they inherit potential biases from the individual tools involved. Nevertheless, this usage is justified as it diminishes the risk of biases that might be present when relying solely on a single modularization tool, thus offering a more reliable inspection regarding potential MR issues.
In the remainder of this section, we first define \emph{PairSmell} and then present an automated approach for its identification.

\subsection{PairSmell Definition}
\label{sec: smell definition}

We define a \emph{PairSmell} as a 3-tuple regarding a pair of entities $e_i$ and $e_j$, where the actual MR violates its apt MR:
\begin{equation}\label{for: smell}
\!\mathit{PairSmell} =\, <(e_i,e_j), \mathit{MR_{act}} (e_i,e_j), \mathit{MR_{apt}} (e_i,e_j)> 
\end{equation}

The first element $(e_i,e_j)$ denotes a pair of entities in a target system, where $e_i \ne e_j $. In this study, we consider an entity as a single code file, following the common practice of architecture-level analyses (\eg~\cite{mo2016decoupling,garcia2021forecasting}).
Both the second and third elements ($\mathit{MR_{act} (e_i, e_j)}$ and $\mathit{MR_{apt} (e_i, e_j)}$) denote modular relations between the entities $e_i$ and $e_j$. The MR of a pair in a specific design $d$ is separated or collocated, formally:
\begin{equation}\label{for: MR}
    \mathit{MR_{d}} (e_i, e_j) =\begin{cases}
    \text{\textit{0}}, & \text{if $mod_d(e_i) \neq mod_d(e_j)$}\\
    \text{\textit{1}}, & \text{if $mod_d(e_i)=mod_d(e_j)$}
    
\end{cases}
\end{equation}
where $mod_d(e_i)$ is the module to which $e_i$ belongs in design $d$.
$\mathit{MR_{act}(e_i, e_j)}$ is the actual modular relation of the pair, which could be extracted from a system snapshot.
Inspired by \emph{consensus clustering}~\cite{fred2005combining,zhang2022weighted}, this work considers an MR apt if it is agreed upon by multiple modularization tools. 
In contrast, if modularization tools disagree, it suggests that the pair may be reasonably designed as either collocated or separated. %In such scenarios, it is deemed that an ideal MR does not exist for the pair.
Formally, an apt MR exists if: 
\begin{equation}
     % \mathit{MR_{apt}(e_i, e_j)} \Leftrightarrow 
     \mathit{MR_{d_1}(e_i, e_j)} = ...= \mathit{MR_{d_m}(e_i, e_j)}
\end{equation}
where $m$ is the number of modularization tools considered.

\subsection{Identification Approach}
\label{sec: smell identification}

\begin{figure}
  \centering
  \includegraphics[width=\linewidth]{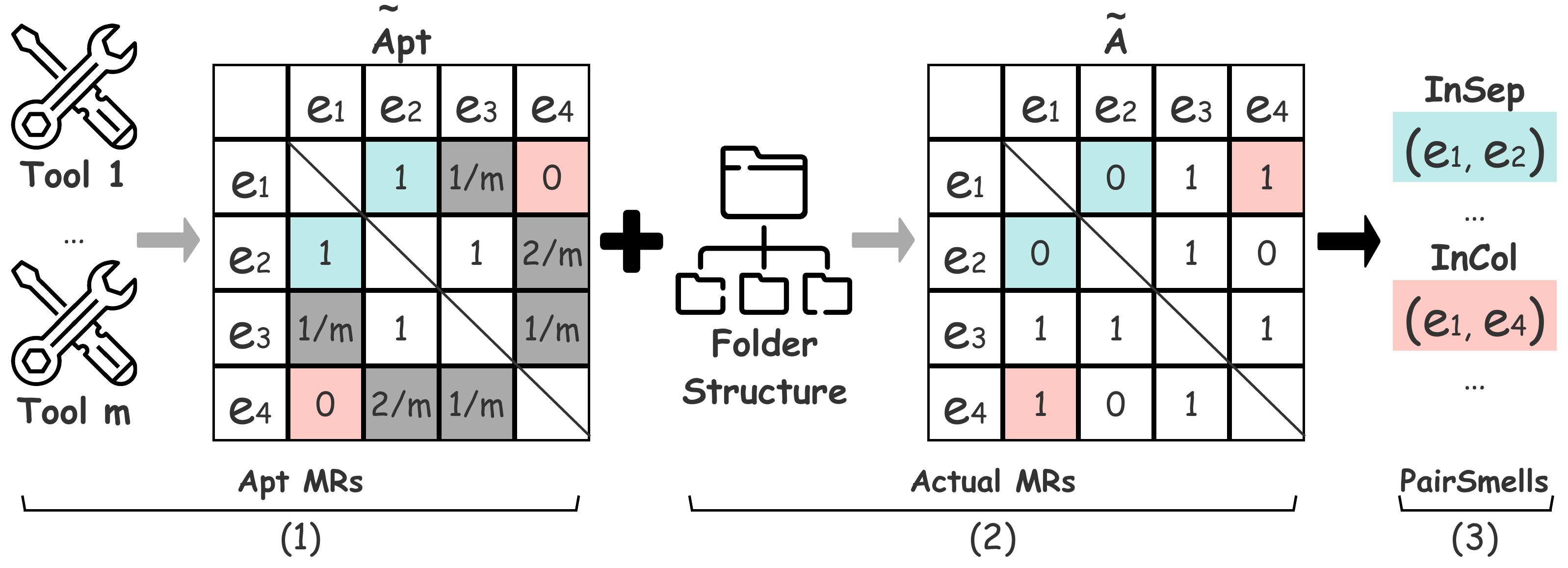}
  \caption{Overview of identifying PairSmell.}
  \label{fig: overview}
\end{figure}

For \emph{PairSmell} identification, we first infer the apt MRs agreed upon by multiple modularization tools, and then utilize them as references to identify smell candidates. 
Fig.~\ref{fig: overview} illustrates our approach in three steps.

\subsubsection{Inferring Apt MRs}

This step infers apt MRs by comparing $m$ solutions from distinct modularization tools.

Given $n$ entities in a system, the solution from modularization tool $t_i$ could be denoted as $\mathit{Mod_{i}} = \left\{mod_i(e_1), ..., mod_i(e_n)\right\}$, where $mod_i(e_j)$ is the module to which $e_j$ belongs in the solution.
We could construct a Co-Association matrix $\Tilde{Apt} \in \mathbb{R} ^{n \times n}$, as defined in the consensus clustering field~\cite{jia2023ensemble,fred2005combining}, to denote the frequency that $e_i$ and $e_j$ occur in the same module across $m$ solutions:
\begin{equation}
    \Tilde{Apt}_{ij} = \frac{1}{m} \sum_{k=1}^{m} \mathit{MR_{k}} (e_i, e_j)
\end{equation}

where $\mathit{MR_{k}} (e_i, e_j)$ is the modular relation between $e_i$ and $e_j$ in the solution from modularization tool $t_k$.

In an $\Tilde{Apt}$ matrix, if most of the modularization tools separate $e_i$ and $e_j$ (\ie $\Tilde{Apt}_{ij}$ near to $0$), these two entities are very likely to belong to different modules. 
Similarly, 
if most of the tools group $e_i$ and $e_j$ into the same module (\ie $\Tilde{Apt}_{ij}$ near to $1$), the two entities are very likely to belong together. 
On the contrary, for the entity pairs with $\Tilde{Apt}_{ij}$ between $0$ and $1$ but near to neither, the modularization tools suggest relatively inconsistent MRs. That is, these pairs could be reasonably implemented as either \textit{separated} or \textit{collocated}, and there is no a promisingly apt MR for them.
We define two types of apt MRs that could be inferred from the matrix.

\begin{itemize}
    \item \textit{apt separated}: denotes a pair which should be separated according to the tools. This occurs when all tools consistently suggest separating the MR for a pair,
    % the suggested MRs for a pair are consistently separated by all tools
    \ie $\Tilde{Apt}_{ij} = 0$. 
    % All such pairs in a system are referred to as $\mathit{AptSep}$.
    
    \item \textit{apt collocated}: indicates a pair which should be collocated. An \textit{apt collocated} exists if the suggested MRs for a pair are collocated by all tools, \ie $\Tilde{Apt}_{ij} = 1$. 
    % All such pairs in a system are denoted by $\mathit{AptCol}$.
\end{itemize}

A cell in the matrix denotes an apt MR if all modularization tools agree with the MR, \ie with $\Tilde{Apt}_{ij}$ as $0$ or $1$ (as Fig.~\ref{fig: overview}). 
Note that we left out those pairs whose MRs are inconsistent (\ie $0 < \Tilde{Apt}_{ij} < 1$), denoted as gray in the figure.
In this sense, we reduce the biases that could be introduced by individual modularization tools (\eg due to specific rules), thereby enhancing the reliability of the apt MRs we derive.

\subsubsection{Recovering Actual MRs}
\label{sec: actual MRs}
In this step, we collect the actual MRs from a system's existing modules.

A key question is, what are the existing `modules' in a system? 
In this study, we consider the folder structure of a system as its existing modules and extract from it the actual MRs. This is because folders represent the actual code organization structure in the development environment, which is created by the developers of the systems~\cite{garcia2021forecasting}.
In fact, folders display a development architectural view, dating back to Kruchten's seminal 4+1 view model~\cite{kruchten19954+}.

Folder structure can be represented by a tree hierarchy of folders and sub-folders. Each leaf of the tree is an entity contained in a folder, which itself may belong to a higher-level folder (super-folder). 
To align with the prevailing notion of mutual exclusive modules in software engineering, \eg in~\cite{sarhan2020software,yang2022enhancing}, we do not consider all folders as `existing' modules. Instead, we select only the lowest level folders to serve as the existing modules for specificity.
Consequently, two entities are considered co-located in the same module only if they reside in exactly the same folder.

Based on the recovered modules, we define an actual MR matrix $\Tilde{A} \in \mathbb{R} ^{n \times n}$, as Fig.~\ref{fig: overview}.
Each cell of $\Tilde{A}$ represents the MR between two entities within the existing system. The possible value for each cell adheres to formula~\ref{for: MR}. 
% Accordingly, all pairs in a system could be divided into two sets, $\mathit{Sep}$ or $\mathit{Col}$, depending on the actual MR value.

\subsubsection{Identifying PairSmell Candidates}

Next, we detect smell candidates, by comparing the apt MRs with the actual MRs. 

For each pair of entities $e_i$ and $e_j$, the apt MR and the actual MR constitute two binary expressions: $\mathit{MR_{apt} (e_i,e_j)}$ and $\mathit{MR_{act} (e_i,e_j)}$. 
The possible value for each expression is \emph{1} or \emph{0}.
We enumerated all 4 combinations of these two expressions. The 2 combinations with consistent $\mathit{MR_{apt} (e_i,e_j)}$ and $\mathit{MR_{act} (e_i,e_j)}$ indicate that the actual MR between $e_i$ and $e_j$ is appropriate, as it aligns with the promisingly apt design. For the other two combinations where $\mathit{MR_{apt} (e_i,e_j)}$ and $\mathit{MR_{act} (e_i,e_j)}$ are inconsistent, we define 2 specific forms of \emph{PairSmell}---\emph{InSep} and \emph{InCol} as follows:

\textbf{Inapt Separated (InSep)}---two entities are separated into different modules in the actual system but the \emph{apt} MR is collocated according to modularization tools. This smell means that the two separated entities are highly related, \eg they may depend on each other, thus all tools group them together. The inapt MR of these two entities may hamper the independence of corresponding modules, making changes of one module propagating to another module~\cite{arvanitou2015introducing}. All entity pairs that match this form is denoted by $\mathit{InSep}$ and identified by: 
\begin{equation}
    \mathit{InSep} = \left\{(e_i,e_j)| \neg \mathit{MR_{act} (e_i,e_j)} \wedge \mathit{MR_{apt} (e_i,e_j)} \right\} 
\end{equation}

\begin{figure}
  \centering
  \includegraphics[width=.8\linewidth]{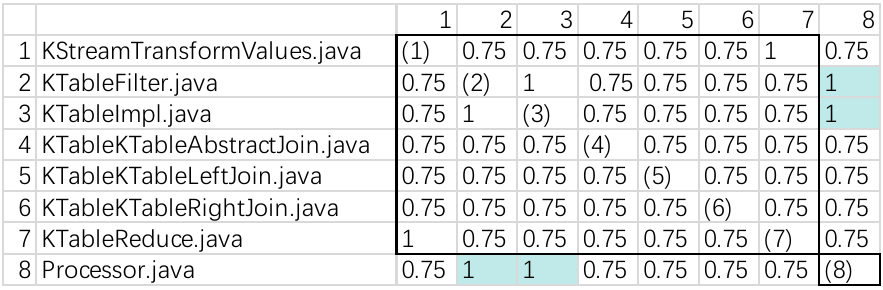}
  \caption{An InSep example. Each number indicates the average frequency that two entities are grouped together by tools. Entities in a lined rectangle actually belong to one module.}
  \label{fig: case_insep}
\end{figure}

Fig.~\ref{fig: case_insep} shows an instance of $\mathit{InSep}$ detected in the project \emph{Kafka}. \emph{Processor} is located in a separate module from all other files. However, we can see from the cells annotated with a number of $1$ that all tools assigned it to be collocated with files \emph{KTableFilter} (row 2) and \emph{KTableImpl} (3).

\begin{figure}
  \centering
  \includegraphics[width=.8\linewidth]{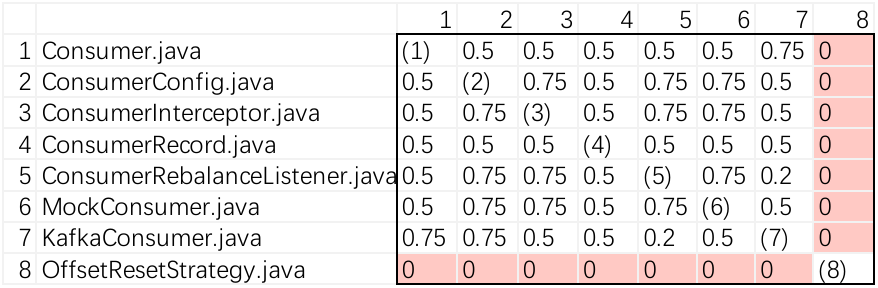}
  \caption{An InCol example with the same annotations as Fig.~\ref{fig: case_insep}.}
  \label{fig: case_incol}
\end{figure}

\textbf{Inapt Collocated (InCol)}---two entities are actually implemented as colloated but the \emph{apt} MR by all tools is to separate them. This smell indicates that the two entities combined in one module are to some extent irrelevant, \eg they address different (or even orthogonal) concerns. This inapt MR may impede the cohesion of the current module, violating the single responsibility principle~\cite{martin2018clean}. All pairs that match this form are denoted by $\mathit{InCol}$ and identified as: 
\begin{equation}
    \mathit{InCol} = \left\{(e_i,e_j)| \mathit{MR_{act} (e_i,e_j)} \wedge \neg \mathit{MR_{apt} (e_i,e_j)} \right\} 
\end{equation}

Fig.~\ref{fig: case_incol} depicts an instance of $\mathit{InCol}$ in \emph{Kafka}. \emph{OffsetResetStrategy} is in the same module with other files. However, the cells annotated with a zero number in the DSM reveal that all modularization tools separate it from other files.

\subsection{Tool Implementation}
\label{sec: tool implementation}

To implement the identification of \emph{PairSmell}, we used four modularization tools as the basis to infer apt MRs.
Our selection criteria are five-fold: (1) The tool is able to modularize C/C++ and Java projects, which are our focus in this study; (2) Its analysis unit is the code file, which is the entity considered in this paper; (3) Its approach should either be established with promising results in prior empirical studies (WCA, LIMBO, and ACDC) or advanced from the latest research published within the last five years (like FCA);
(4) Its source code should be accessible; (5) The tool uses deterministic techniques, \ie each execution yields the same result, to avoid the effects of randomness in our results.
The final tools are:

\begin{itemize}
    \item \textbf{WCA}~\cite{maqbool2004weighted} is a hierarchical clustering algorithm using inter-cluster distance to extract software modules. The distance can be calculated by two similarity metrics: \textit{UE} and \textit{UENM}. We used UENM as it outperforms UE~\cite{garcia2013comparative}.

    \item \textbf{LIMBO}~\cite{andritsos2004limbo} is a hierarchical algorithm that clusters categorical data using a distance measure (called mutual information) to minimize information loss in the clusters.

    \item \textbf{ACDC}~\cite{tzerpos2000accd} clusters software entities based on specific patterns (\eg body-header) and uses orphan adoption technique to assign remaining entities to clusters.

    \item \textbf{FCA}~\cite{teymourian2020fast} is a clustering algorithm that maximizes intra-connectivity and minimizes inter-connectivity in clusters. 
\end{itemize}

For running these tools, we provide a dependency graph, as required by the tools. The nodes of the graph represent software entities, and the edges represent the structural relations between entities. 
This paper uses \emph{Depends}~\cite{depends2022} to recover structural relations, as it is capable of extracting 13 dependency types by analyzing the syntactic structures of C/C++/Java programs, such as \emph{call}, \emph{contain}, and \emph{implement}. In this sense, the extensive data collected enables us to recover more accurate dependency graphs of software systems, which generally results in enhanced modularization solutions.

\noindent
\textbf{Tool Evaluation.}
\emph{PairSmell} is defined based on the deviations of actual MRs from the apt MRs. Since there is no a set of \emph{ground-truth} apt MRs, it is hard to construct a validation set. Thus, we decide to manually examine the detection results of our tool, similar to the methodology used by Kim et al.~\cite{kim2021disabled}.

We start by executing the tool on 20 projects, listed in Section~\ref{sec: data collection}, which results in 9,415 smell instances. Then, we derive a sample to be manually validated. We randomly select 370 out of 9,415 smells, based on a 95\% confidence level and 5\% confidence interval~\cite{boslaugh2012statistics}. This includes 129 $\mathit{InSep}$ and 241 $\mathit{InCol}$ pairs. Each pair is then independently examined by two authors to decide the correctness. 
Both annotators produced identical results after completing the tagging, and our validation process achieved a precision of 100\%.
However, we cannot evaluate the recall due to the lack of oracles.

\section{Empirical Study}

The \emph{goal} of this study is to provide empirical evidence for assessing the severity of \emph{PairSmell} in practice, focusing particularly on its prevalence, impact, and evolution—three fundamental aspects critical to investigating a phenomenon~\cite{muse2020prevalence,liu2024prevalence,mendes2022dazed,khan2022automatic}. Specifically, the study addresses three research questions:

\begin{enumerate}[label={RQ\arabic*.}]
    \item \textbf{\textit{To what extent does PairSmell appear in software projects?}}
    This question aims to quantitatively assess the prevalence of \emph{PairSmell} in software projects. If \emph{PairSmell} is prevalent, \aka its amount is notable per project, it suggests that the proposed smell merits further attention.

    \item \textbf{\textit{To what extent does PairSmell impact software maintenance?}} 
    This question assesses how \emph{PairSmell} affects software maintenance by analyzing the co-change extent manifested in project revision history. If the co-change extent of smelly pairs significantly and detrimentally differs from non-smelly pairs, it indicates a deviation from the ideal modular structure of well-maintained systems.

    \item \textbf{\textit{How does the amount of PairSmells evolve across time?}} 
    With this question, we aim to investigate the amount of \emph{PairSmell} as systems evolve. 
    If \emph{PairSmell} proliferates, or at least does not diminish, across time, it would denote a significant motivation for its removal.

\end{enumerate}

\subsection{Data Collection}
\label{sec: data collection}

For empirically answering the research questions, we choose open-source software projects as study subjects by following three predefined criteria:
(1) C/C++ and Java projects on GitHub because they are among the most popular programming languages;
% they can be analyzed automatically by \emph{Depends} (cf. Section~\ref{sec: tool implementation});
(2) projects with at least 2 years of change history and over 1,000 commits, so that they can provide sufficient evolution data for analyzing the impact of \emph{PairSmell} in software evolution and maintenance;
(3) non-trivial projects with at least 100 entities, because architecture smells turn to be significant especially for non-trivial projects~\cite{liu2024prevalence}.
The selected projects are shown in Table~\ref{tab: systems}, together with their number of entities (\emph{\#Entity}), relationships between the entities (\emph{\#Link}), and commits (\emph{\#Cmt}). These projects differ in their scale, business domains, and other characteristics. All data we used are publicly available~\cite{replication}.

\begin{table}
\caption{Summary of the studied software projects.}
\label{tab: systems}

\scriptsize
\begin{tabular}{p{6pt}p{25pt}p{64pt}p{18pt}rrr}
\toprule
\textbf{$P_i$} & \textbf{Project(l)} & \textbf{Domain} & \textbf{Version} & \textbf{\#Entity} & \textbf{\#Link} & \textbf{\#Cmt} \\
\midrule
$P_1$ & Arrow(c) & Memory analytics & 0.15.0 & 568 & 3,003 & 5,159 \\
$P_2$ &Brpc(c) & RPC framework & 1.5.0 & 385 & 345 & 2,032 \\
$P_3$ &Cassandra(j) & Row store & 0.6.10 & 283 & 5,569 & 1,752 \\
$P_4$ &Druid(j) & Analytics database & 0.7.0 & 1,045 & 7,651 & 4,980 \\
$P_5$ &Gobblin(j) & Data management & 0.9.0 & 1,279 & 9,743 & 3,717 \\
$P_6$ &Hadoop(j) & Distributed framework & 0.20.0 & 890 & 17,266 & 3,461 \\
$P_7$ &Hbase(j) & Storage system & 1.0.2 & 1,456 & 34,968 & 10,061 \\
$P_8$ &Httpd(c) & Web server & 2.0.46 & 229 & 3,349 & 11,539 \\
$P_9$ &Impala(c) & SQL framework & 2.7.0 & 439 & 490 & 4,934 \\
$P_{10}$ &Iotdb(j) & Data management & 0.11.0 & 836 & 19,273 & 4,209 \\
$P_{11}$ &Kafka(j) & Event streaming & 0.10.2.1 & 747 & 11,593 & 3,247 \\
$P_{12}$ &Kudu(c) & Storage engine & 0.7.0 & 514 & 78 & 4,022 \\
$P_{13}$ &Kvrocks(c) & NoSQL database & 2.8.0 & 220 & 4,716 & 1,262 \\
$P_{14}$ &Lucene(j) & Searchh engine & 2.9.2 & 1,006 & 21,377 & 4,042 \\
$P_{15}$ &Mahout(j) & DSL framework & 0.6 & 1,052 & 12,939 & 2,269 \\
$P_{16}$ &Mesos(c) & Cluster manager & 0.21.2 & 281 & 554 & 3,713 \\
$P_{17}$ &Ozone(j) & Object store & 1.0.0 & 1,380 & 8,595 & 2,698 \\
$P_{18}$ &Pulsar(j) & Pub-sub messaging & 2.3.0 & 1,142 & 20,519 & 2,892 \\
$P_{19}$ &Thrift(c) & RPC framework & 0.12.0 & 202 & 483 & 5,384 \\
$P_{20}$ &Traffic(c) & Caching proxy server & 4.2.0 & 963 & 34,589 & 4,301\\
\bottomrule
\end{tabular}

\end{table}

\subsection{RQ1: Prevalence of PairSemll}
\label{sec: prevalence}

\subsubsection{Setup}
To answer RQ1, we identify \emph{PairSmell} on the current version of 20 projects (cf. Table~\ref{tab: systems}). We study how frequently \emph{PairSmell} appears at both pair and entity levels. 
Please note that an entity affected by \emph{PairSmell} is involved in at least one smell instance. 
To provide a comparative statistic, we calculate the percentages of \emph{PairSmell} relative to the total number of corresponding program elements (pairs or entities). For example, we calculate at the pair level, the proportion of $\mathit{InSep}$ among all separated pairs in a project, indicating the extent of inappropriate MR design for separated pairs. In addition, we calculate smell density to measure the `smelliness' of a specific smell form $x$ ($\mathit{InSep}$ or $\mathit{InCol}$) among the affected entities. This metric quantifies the average number of smell instances concurrently affecting each entity, and is computed as follows: 
\begin{equation}
    \mathit{Density (x)} = \frac{\mathit{Total}\; \mathit{instances}\; \mathit{of}\; x \times 2}{\mathit{Number} \; \mathit{of} \; \mathit{entities} \; \mathit{involved}}
\end{equation}

\subsubsection{Results}

\begin{table}
\caption{InSep and InCol instances in the current version. }
\label{tab: smell proportion}
\scriptsize
\centering
\begin{tabular}{p{6pt}rrr r@{\hspace{0.001cm}} rrr}
\toprule
 & \multicolumn{3}{c}{\textbf{InSep}} &  &\multicolumn{3}{c}{\textbf{InCol}} \\
 \cmidrule{2-4} \cmidrule{6-8}
\multirow{-2}{*}{\textbf{$P_i$}} & \textbf{Pair(\%)} & \textbf{Entity(\%)} & \textbf{Density} &  & \textbf{Pair(\%)} & \textbf{Entity(\%)} & \textbf{Density} \\
\midrule
{$P_1$} & 3($<$0.01) & 6(1.06) & 1.00 & & 143(1.98) & 86(15.14) & 3.33  \\
{$P_2$} & 0(0)& 0(0) & 0.00 && 13(0.29) & 13(3.38) & 2.00 \\
{$P_3$} & 27(0.07)	& 34(20.14) &1.59  && 365(16.60) & 171(60.42)&4.27  \\
{$P_4$} & 85(0.02)	& 119(11.39) &	1.43  & &90(1.01)	&63(5.00)	&2.86  \\
{$P_5$} & 80(0.01)	&129(10.09)&	1.24  && 61(0.96)	&64(5.00)	&1.91  \\
{$P_6$} & 334(0.09)&	242(27.19)	&2.76  && 162(1.08)	&114(12.81)&	2.84 \\
{$P_7$} & 960(0.09)&	434(29.81)	&4.42  && 176(0.73)	&65(4.46)&	5.42  \\
{$P_8$} & 1($<$0.01)&	2(0.87)&	1.00  && 199(15.05)	&95(41.49)	&4.19  \\
{$P_{9}$} & 0(0)&	0(0)	&0.00  & &57(0.51)&	43(9.80)	&2.65  \\
{$P_{10}$} & 134(0.04)&	167(19.98)	&1.60  && 189(5.90)	&172(20.57)&	2.20 \\
{$P_{11}$} & 93(0.04)&	125(16.73)&	1.49  & &66(0.68)&	80(10.71)	&1.65 \\
{$P_{12}$} & 1($<$0.01)&	2(0.39)	&1.00 && 36(0.27)	&26(5.06)&	2.77 \\
{$P_{13}$} & 0(0)&	0(0)&	0.00 && 335(14.56)	&130(59.09)	&5.15  \\
{$P_{14}$} & 398(0.08)	&279(27.73)&	2.85 & &536(2.70)	&332(33.00)&	3.23  \\
{$P_{15}$} & 987(0.18)&	349(33.18)&	5.66  && 36(0.73)	&39(3.71)	&1.85  \\
{$P_{16}$} & 0(0)	&0(0)&	0.00 && 75(2.19)	&33(11.75)&	4.55  \\
{$P_{17}$} & 119(0.01)	&151(10.94)&	1.58  && 44(0.65)	&48(3.48)&	1.83 \\
{$P_{18}$} & 66(0.01)	&108(9.46)&	1.22  && 113(1.33)	&114(9.98)	&1.98 \\
{$P_{19}$} & 0(0)	&0(0)	&0.00  && 67(5.64)	&32(15.84)	&4.19  \\
{$P_{20}$} & 0(0)&	0(0)	&0.00  & &3,364(16.10)	&742(77.05)&	9.07 \\
\midrule
{\textbf{Avg.}} & 164(0.03)	&107(14.60)&	1.44  && 306(4.45)	&123(20.44)	&3.40 \\
\bottomrule
\end{tabular}
\end{table}

Table~\ref{tab: smell proportion} presents the prevalence of $\mathit{InSep}$ and $\mathit{InCol}$ in different projects. The 2nd and 5th column show the numbers and percentages of $\mathit{InSep}$ and $\mathit{InCol}$ at pair level.
%Results show that, the number of $\mathit{InSep}$ instances is over 60 in more than half of the projects.
On average, 164 $\mathit{InSep}$ pairs were identified in each project. 
For $\mathit{InCol}$, the average number of smells could be as high as 306 (over 4\%). 
In certain projects, \eg $P_2$, only a few \emph{PairSmells} were identified, indicating that the MR design in these projects tends to be structurally sound. Overall, the presence of \emph{PairSmell} is noteworthy across the 20 projects.

From the 3rd and 6th column, both $\mathit{InSep}$ and $\mathit{InCol}$ are widespread among software entities in the projects. 
About 15\% of entities in each project are affected by $\mathit{InSep}$, while a higher average is observed for $\mathit{InCol}$.
%, the value is high at 20\%, and 55\% projects have over 10\% entities being affected.
That is, a substantial proportion of entities are impacted by \emph{PairSmell} in these projects.

Columns 4 and 7 present the smell \emph{density} among affected entities.
Results show that each `smelly' entity is involved, on average, in 1.44 $\mathit{InSep}$ pairs, and 3.40 $\mathit{InCol}$ pairs. 
%This implies that an entity impacted by $\mathit{InCol}$ is generally (2.36x) more smelly than an entity affected by $\mathit{InSep}$.
% , since the former is involved in 2.36 times more smell instances.

\begin{figure}[b]
  \centering
  \includegraphics[width=0.84\linewidth]{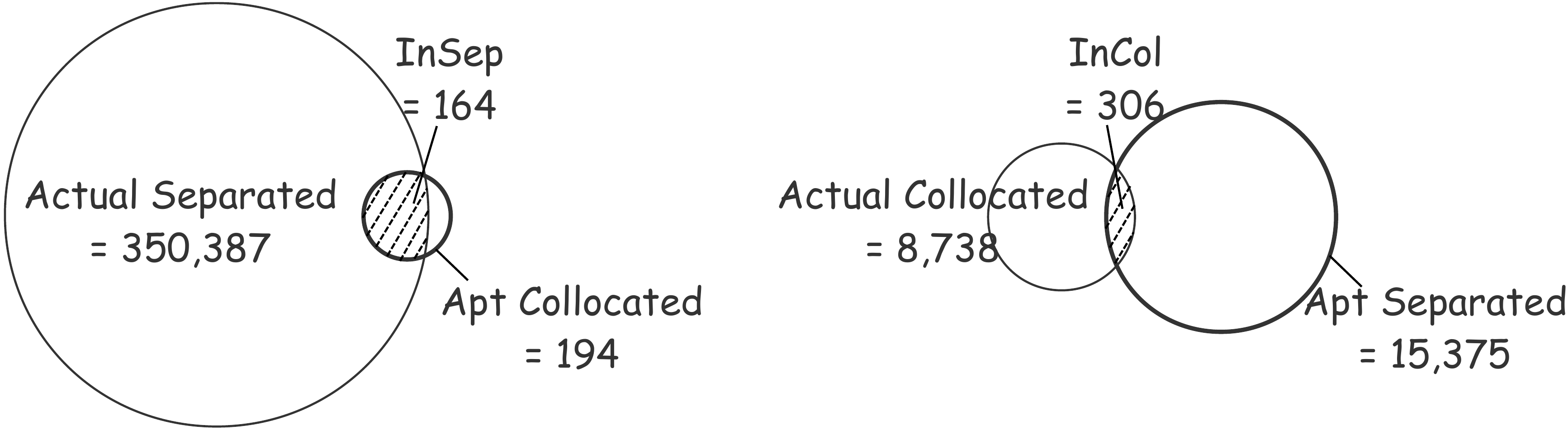}
  \caption{The sets of InSep and InCol, averaged over 20 projects.}
  \label{fig: insep_incol_set}
\end{figure}

To explore the differences between $\mathit{InSep}$ and $\mathit{InCol}$, Fig.~\ref{fig: insep_incol_set} shows the number of separated and collocated pairs aggregated across all projects, and how they overlap with the apt MRs. The two overlapping parts constitute the sets of $\mathit{InSep}$ and $\mathit{InCol}$ `smelly' pairs respectively. 
For example, among the 194 pairs where all modularization tools design them to be collocated (\ie apt collocated), 164 (84.5\%) are actually implemented as separated (\ie $\mathit{InSep}$). 
Such structuring into different modules increases inter-module coupling.
In contrast, only 2.0\% apt separated pairs are actually implemented as collocated (306 out of 15,375), suggesting developers' caution for structuring responsibilities into modules.

% Finally, we notice that the number of actually collocated pairs is significantly lower than separated pairs (8,738 vs. 350,387). This observation explains why the percentage of $\mathit{InSep}$ pairs is much lower than that of $\mathit{InCol}$ in Table~\ref{tab: smell proportion}.

\begin{tcolorbox}[boxsep=0pt,left=2pt,right=2pt,top=3pt,bottom=2pt,arc = 0.1mm, boxrule = 0.3mm] \itshape
\textbf{RQ1 Summary:} 
Both $\mathit{InSep}$ and $\mathit{InCol}$ are prevalent in the dataset. %, with 164 and 306 instances on average.
Developers seem more inclined to organize highly relevant entities into separate modules (i.e., incur $\mathit{InSep}$), thus introducing inter-module coupling, than grouping responsibilities within the same module (i.e., incur $\mathit{InCol}$).
\end{tcolorbox}

\subsection{RQ2: Impact on Software Maintenance}
\label{sec: impacts}

Code revision history, is frequently used as a benchmark to investigate the impact of generic smells, \eg how smells impact fault- and change-proneness~\cite{khomh2009exploratory,xiao2021detecting}, smells' impact on maintainability~\cite{sjoberg2012quantifying,yamashita2012code,jin2023dependency} or file co-change~\cite{mo2023exploring}. 
Given that \emph{PairSmell} describes a problematic relationship between entities, our evaluation focuses on its impacts on file co-change relation.
The underlying principle is, within a healthy modular structure, files in the same module should change together, while files from different modules should change independently.
This RQ compares the co-change of smelly versus non-smelly pairs, within and between modules, to explore if \emph{PairSmell} disrupts the expected healthy structure.

\setcounter{figure}{8}

\begin{figure*}[b]
    \centering
    \subfloat[The values of $\mathit{K_{COR}}$ for InSep.]{\includegraphics[width=.32\textwidth]{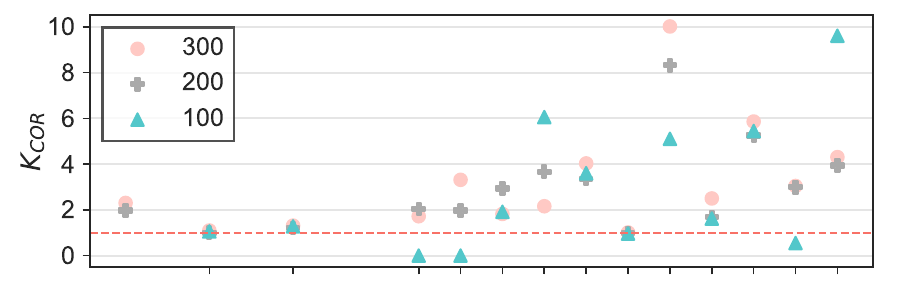}%\caption{fig1}
    }
    % \quad
    \subfloat[The values of $\mathit{K_{CCO}}$ for InSep.]{\includegraphics[width=.32\textwidth]{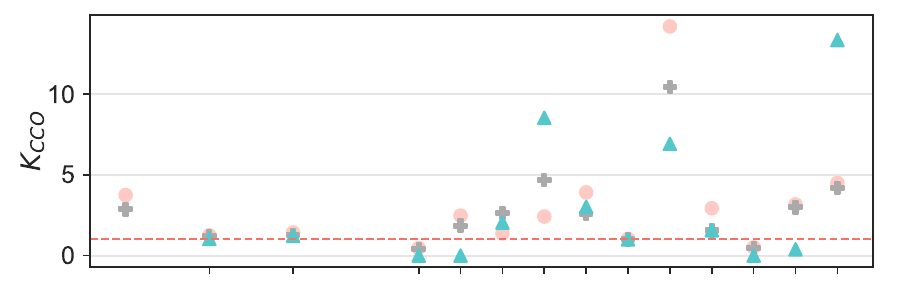}%\caption{fig1}
    }
    % \quad
    \subfloat[The values of $\mathit{K_{DOR}}$ for InSep.]{\includegraphics[width=.32\textwidth]{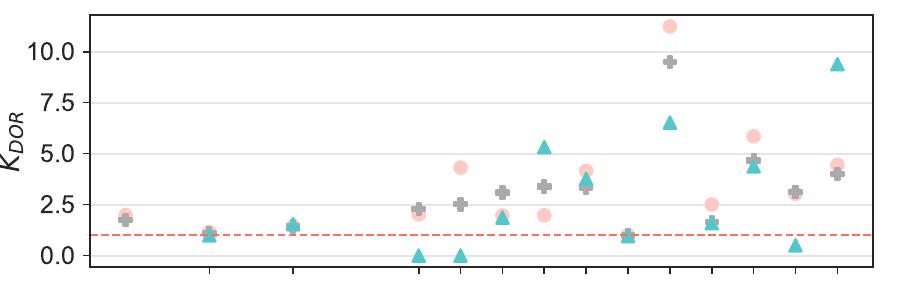}%\caption{fig1}
    }
    \quad
    \subfloat[The values of $\mathit{K_{COR}}$ for InCol.]{\includegraphics[width=.32\textwidth]{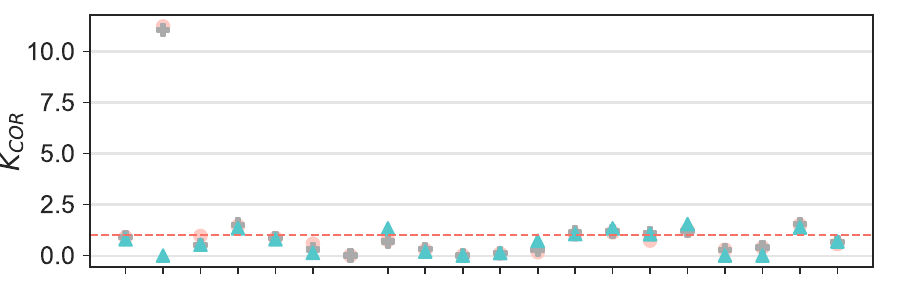}%\caption{fig1}
    }
    % \quad
    \subfloat[The values of $\mathit{K_{CCO}}$ for InCol.]{\includegraphics[width=.32\textwidth]{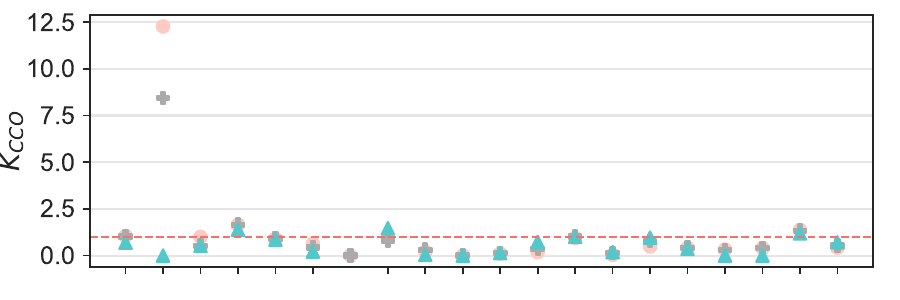}%\caption{fig1}
    }
    % \quad
    \subfloat[The values of $\mathit{K_{DOR}}$ for InCol.]{\includegraphics[width=.32\textwidth]{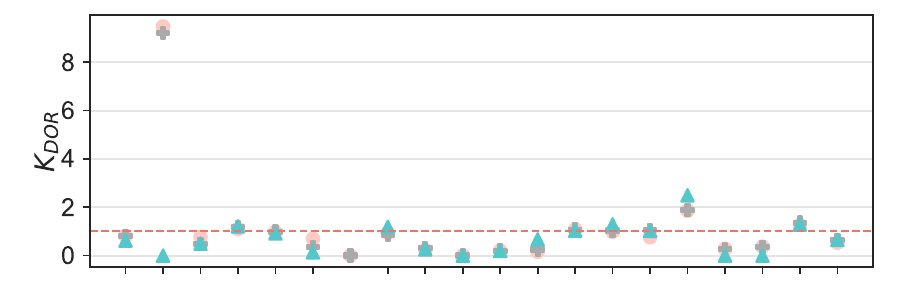}%\caption{fig1}
    }
    \quad
    \caption{The distributions of $\mathit{K_{COR}}$, $\mathit{K_{CCO}}$, $\mathit{K_{DOR}}$ for InSep and InCol in all projects. Red lines ``\protect\redtext{}{- - -}'' mean $\mathit{K_{mtr}=1}$.} 
    \label{fig: distributions}
    \vspace*{-3.0ex}
\end{figure*}

\setcounter{figure}{7}

\begin{figure}
  \centering
  \includegraphics[width=0.86\linewidth]{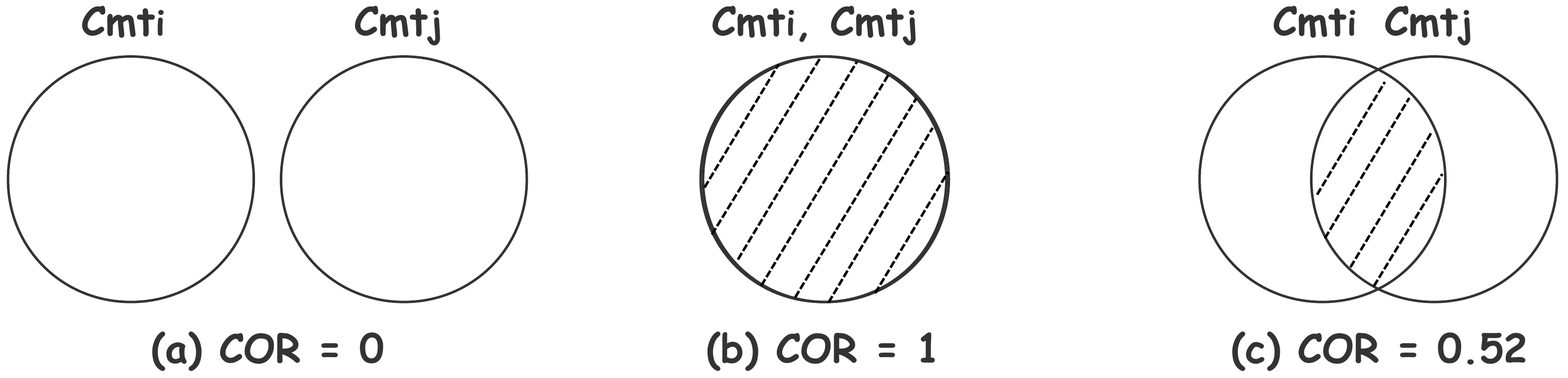}
  \caption{The commit sets of two entities overlap differently.}
  \label{fig: commit sets}
\end{figure}

\setcounter{figure}{9}

\subsubsection{Setup}
File co-change in prior studies is typically captured by the absolute frequency, \ie the number of commits that a file pair change together~\cite{jin2020exploring,mo2023exploring}, which is inefficient for comparing co-change extents among different pairs~\cite{chambers2014inefficiency}.
Additionally, various types of evidence, not just commits, have been used to measure software maintenance and changes~\cite{xiao2021detecting,chowdhury2022revisiting}.
To robustly assess \emph{PairSmell}'s impact, we propose a suite of measures based on relative measurement theory~\cite{allen2001introduction,zhong2023measuring,zimmermann2005mining}, utilizing commonly used evidence, as illustrated in Fig.~\ref{fig: commit sets}.

For any two entities $e_i$ and $e_j$, their commit sets~\cite{jin2020exploring} during a specific time period can be represented as $\mathit{Cmt_i}$ and $\mathit{Cmt_j}$. If two entities changed in completely different commits, as Fig.~\ref{fig: commit sets} (a), they are likely independent and can change without affecting each other. On the contrary, if $e_i$ and $e_j$ shared exactly the same commit sets, as Fig.~\ref{fig: commit sets} (b), these entities co-changed consistently. Based on these observations, we define three measures to quantify the extent of co-change between a pair:

\noindent
\textbf{\textit{1. Commit Overlap Rate (COR):}} measures the extent changes made to two entities overlap. $\mathit{COR = \frac{2*|Cmt_{i,j}|}{|Cmt_i|+|Cmt_j|}}$, where $\mathit{Cmt_i}$ is the commit set that changed entity $e_i$, $\mathit{Cmt_{i,j}}$ is the commit set where entities $e_i$ and $e_j$ changed together. A larger $\mathit{COR}$ means more overlap between two entities' commits, indicating that these entities are more relevant.

\noindent
\textbf{\textit{2. Code Change Overlap (CCO):}} measures the likelihood that code changes~\cite{Mo2015,mo2019architecture} to two entities occurred simultaneously. 
$\mathit{CCO = \frac{|Ch_{i,j}|}{|Ch_i|+|Ch_j|}}$, where $Ch_i$ is the lines of code changed in entity $e_i$, $\mathit{Ch_{i,j}}$ is the lines of code changed in either $e_i$ or $e_j$ that occurred together in the same commit. 
$\mathit{Ch_{i,j}}$ is counted once because $\mathit{Ch_i}$ and $\mathit{Ch_j}$ do not intersect. 
The larger the $\mathit{CCO}$, the more often two entities undergo simultaneous code changes, indicating a more relevant pair. 

\noindent
\textbf{\textit{3. Developer Overlap Rate (DOR):}} measures to what extent the sets of developers~\cite{mo2016decoupling,jin2020exploring} changing two entities overlap.
$\mathit{DOR = \frac{2*|Dev_{i,j}|}{|Dev_i|+|Dev_j|}}$, where $Dev_i$ indicates the developer set changing entity $e_i$, $Dev_{i,j}$ is the developer set changing both $e_i$ and $e_j$. The higher the value, the more likely the two entities were changed by the same developers, suggesting a possibly greater relevance between them.

We use $\mathit{K_{COR}}$, $\mathit{K_{CCO}}$, and $\mathit{K_{DOR}}$ to comprehensively assess the relative co-change of a smelly pair as compared with that of a non-smelly pair, similar to the work of Mo et al.~\cite{mo2019architecture}. 
Our hypothesis is that an $\mathit{InSep}$ pair is more likely to be related than other separated pairs, and thus more co-changed; in contrast, an $\mathit{InCol}$ pair is less likely to be related than other collocated pairs and therefore less co-changed.
The detailed measures are as follows:
\begin{equation}
\label{for: k_values}
    \mathit{K_{mtr}} = \frac{\mathit{mtr}\; \mathit{of}\; \mathit{Smelly}\;\mathit{pairs} \;\mathit{(avg.)}}{\mathit{mtr}\; \mathit{of}\; \mathit{Non\!-\!Smelly}\; \mathit{pairs} \;  \mathit{(avg.)}}\\
\end{equation}
% \begin{equation}
% \label{for: insep}
%     \mathit{K_{mtr}} = \frac{\mathit{mtr}\; \mathit{of}\; \mathit{pairs} \;\mathit{in} \;\mathit{InSep} \;\mathit{(avg.)}}{\mathit{mtr}\; \mathit{of}\; \mathit{pairs}\; \mathit{in}\; \mathit{Separated - InSep} \;  \mathit{(avg.)}}\\
% \end{equation}
% \begin{equation}
% \label{for: incol}
%    \mathit{K_{mtr}} = \frac{\mathit{mtr}\; \mathit{of}\; \mathit{pairs}\; \mathit{in}\; \mathit{InCol} \; \mathit{(avg.)}}{\mathit{mtr}\; \mathit{of}\; \mathit{pairs} \; \mathit{in} \; \mathit{Collocated -InCol} \; \mathit{(avg.)}}
% \end{equation}

\noindent
where $\mathit{mtr}$ can be $\mathit{COR}$, $\mathit{CCO}$, and $\mathit{DOR}$. For $InSep$, $\mathit{Smelly}$ pairs denote pairs in the $InSep$ set, and $\mathit{Non\!-\!Smelly}$ pairs are those in the set of $\mathit{Separated\! - \!InSep}$. For $InCol$, these are the sets of $InCol$ and $\mathit{Collocated \!-\!InCol}$. For a project, a $\mathit{K_{COR}}$ value (or $\mathit{K_{CCO}}$, $\mathit{K_{DOR}}$) exceeding~1 means that $\mathit{InSep}$ pairs co-changed more frequently than other separated pairs.
Conversely, a value less than 1 suggests that $\mathit{InCol}$ pairs co-changed less frequently than other collocated pairs.

\subsubsection{Results}

Fig.~\ref{fig: distributions} shows the values of $\mathit{K_{COR}}$, $\mathit{K_{CCO}}$, $\mathit{K_{DOR}}$ regarding $\mathit{InSep}$ and $\mathit{InCol}$. These values were calculated by mining 100, 200, and 300 commits before the current version ($\mathit{Delta}$) of each project, to ensure an evaluation with a sufficient evolution history~\cite{xiao2021detecting}. 
We did not mine a project's revision history from its beginning since an identified smell might not be smelly in the initial stages.
Each point in the figure denotes the result for a single project. Some projects have no points in specific analyses, because the corresponding smell sets are empty (as shown for 6 projects in Table~\ref{tab: smell proportion}) or no smelly pairs were changed during the analyzed commits.

Considering the $\mathit{K_{COR}}$ score for $\mathit{InSep}$ as Fig.~\ref{fig: distributions} (a), most of their values are greater than 1, except for 3 values below $\mathit{K_{COR}} = 1$ line.
Similar results can be observed from other scores.
This indicates that, although belonging to different modules, $\mathit{InSep}$ pairs are more likely to be changed together than other separated pairs. 
\emph{Effect size}~\cite{rosenthal1994parametric} results (cf. Table~\ref{tab: test}) show that significant differences (as per T-test~\cite{kim2015t}) are medium to large for most deltas.
As for the $\mathit{K_{COR}}$ values for $\mathit{InCol}$, 14 out of 20 (70\%) values are less than 1 (analyzed using 300 commits). Similar results are observed for other metrics, indicating that despite collocation, $\mathit{InCol}$ pairs are less co-changed in their evolution than other collocated pairs, possibly suggesting a responsibility overload in the modules. Interestingly, the differences are significant only in the analysis using 100 commits but not in that with longer history length. This could be attributed to the variability of $\mathit{InCol}$ smells across time (cf. Section~\ref{sec: evolution}), implying that some $\mathit{InCol}$ instances might not be smelly in a previous version.

\begin{figure}[b]
    \centering
    {\includegraphics[width=.18\textwidth]{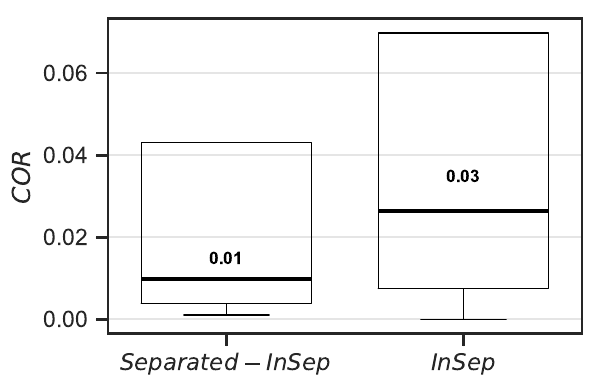}%\caption{fig1}
    }
    {\includegraphics[width=.18\textwidth]{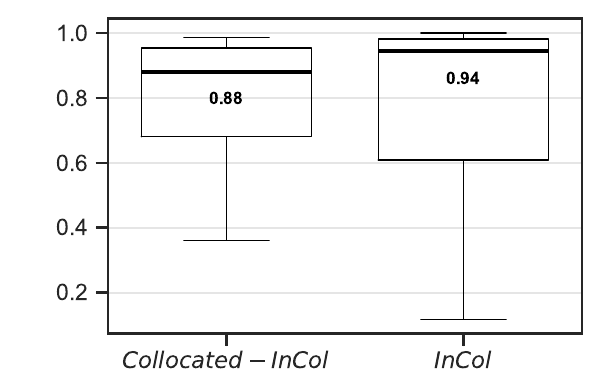}%\caption{fig1}
    }
    \quad
    \caption{$\mathit{COR}$ distribution of different pairs (100 commits).} 
    \label{fig: insep_incol_co_changes}
\end{figure}

On average, the differences of $K$ values for $\mathit{InSep}$ are larger than that for $\mathit{InCol}$.
In the analysis using 100 commits, the average $K$ values for $\mathit{InSep}$ are $\mathit{K_{COR}} = 2.86$, $\mathit{K_{CCO}} = 3.00$, and $\mathit{K_{DOR}} = 2.83$ across all projects. That is, a $\mathit{InSep}$ pair is on average $\frac{(2.86-1)+(3.00-1)+(2.83-1)}{3} = 190\%$ more likely to co-change than a separated pair without smell.
For $\mathit{InCol}$, the averaged values are $\mathit{K_{COR}} = 0.68$, $\mathit{K_{CCO}} = 0.55$, and $\mathit{K_{DOR}} = 0.71$. The likelihood of $\mathit{InCol}$ pairs co-changing is $\frac{(1-0.68)+(1-0.55)+(1-0.71)}{3} = 35\%$ lower than that of other collocated pairs. 
We assume that the difference between $\mathit{InSep}$ and $\mathit{InCol}$ stems from the fact that separated pairs are generally rarely (if ever) modified simultaneously; as a result, the frequent co-changes among $\mathit{InSep}$ pairs appear more evident and detrimental by comparison. In fact, in over 50\% projects, the average $\mathit{COR}$ values of other separated pairs (\ie $\mathit{Separated-InSep}$ in Fig.~\ref{fig: insep_incol_co_changes}) are close to 0.

\begin{table}
\caption{Co-change differences between smelly and non-smelly pairs. \colorbox{mygray}{Gray} results are \textit{significant} differences with $p < .05$.}
\label{tab: test}
\centering
\scriptsize
\begin{tabular}{llllr@{\hspace{0.001cm}}lll}
\toprule
 &  \multicolumn{3}{c}{\textbf{InSep}} & & \multicolumn{3}{c}{\textbf{InCol}} \\
 \cmidrule{2-4} \cmidrule{6-8}
\multirow{-2}{*}{\textbf{Delta}} & $\mathit{K_{COR}}$ & $\mathit{K_{CCO}}$ & $\mathit{K_{DOR}}$ & &$\mathit{K_{COR}}$ & $\mathit{K_{CCO}}$ & $\mathit{K_{DOR}}$ \\
\midrule
% 300 & $p$ & \cellcolor[HTML]{D9D9D9}.002 & \cellcolor[HTML]{D9D9D9}.020 &  \cellcolor[HTML]{D9D9D9}.003 & &.646 & .610 & .607 \\
 300 & \cellcolor[HTML]{D9D9D9}.91 & \cellcolor[HTML]{D9D9D9}.61 & \cellcolor[HTML]{D9D9D9}.88 && .09 & .06 & .06 \\
% 200 & $p$ & \cellcolor[HTML]{D9D9D9}.001 & \cellcolor[HTML]{D9D9D9}.013 & \cellcolor[HTML]{D9D9D9}.002 && .646 & .476 & .601 \\
 200 & \cellcolor[HTML]{D9D9D9}.97 & \cellcolor[HTML]{D9D9D9}.67 & \cellcolor[HTML]{D9D9D9}.95 && .08 & -.01 & .06 \\
% 100 & $p$ & \cellcolor[HTML]{D9D9D9}.020 & .051 & \cellcolor[HTML]{D9D9D9}.020 & &\cellcolor[HTML]{D9D9D9}.011 & \cellcolor[HTML]{D9D9D9}.001 & \cellcolor[HTML]{D9D9D9}.032 \\
 100 & \cellcolor[HTML]{D9D9D9}.64 & .49 & \cellcolor[HTML]{D9D9D9}.64 && \cellcolor[HTML]{D9D9D9}-.57 & \cellcolor[HTML]{D9D9D9}-.91 & \cellcolor[HTML]{D9D9D9}-.45 \\
\bottomrule
\end{tabular}
\end{table}

\begin{tcolorbox}[boxsep=0pt,left=2pt,right=2pt,top=3pt,bottom=2pt,arc = 0.1mm, boxrule = 0.3mm] \itshape
\textbf{RQ2 Summary:} The pairs identified as $\mathit{InSep}$ are 190\% more likely to co-change compared to separated pairs without smells, whereas $\mathit{InCol}$ pairs exhibit 35\% less co-change than proper collocated pairs. Both of these observations indicate that the modular structure is significantly undermined, and software maintenance is adversely affected.
\end{tcolorbox}

\subsection{RQ3: Evolution of PairSmell}
\label{sec: evolution}

In this question, we analyze how the amount of smells changes across time to explore whether \emph{PairSmell} will proliferate in a system if left unaddressed.

\subsubsection{Setup} To answer RQ3, we gather all smell instances for each project across its evolution history.
Considering that each commit may alter the architecture and affect the smell instances, it would be strenuous to analyze each commit in the history. Instead, we opt to analyze snapshots by selecting one commit every two weeks before the current version in Table~\ref{tab: systems}.
Our goal is to capture the evolution activities over approximately a year, which results in 25 snapshots for each project (including the current version).
However, some projects may not experience changes during certain periods; therefore our analysis ultimately covers a total of 473 distinct snapshots.

To conduct a global analysis of $\mathit{InSep}$ and $\mathit{InCol}$, we aggregate the percentages of smells at both the pair and entity levels across all projects and then compute the average values. We choose not to analyze the absolute number of smells, as the increase of this value could be attributed to the growing system size according to prior studies~\cite{gil2017correlation,soto2021longitudinal}.
We represent the average percentages at each level as a time series: $s_1, ..., s_{25}$, where $s_i$ is the averaged percentage for that level across all projects at the $i$-th snapshot. We collect time series for $\mathit{InSep}$ and $\mathit{InCol}$ respectively.

For each smell form, we determine the overall evolution trend for the percentage of smells: increase, decrease, or stable.
We notice a non-monotonic trend in the percentage of smells, \ie the value increases and decreases at different time intervals. To account for such a non-monotonic trend, we fit a simple linear regression model, denoted as $lm$, and determine the trend by examining the sign of the $slope$ of the regression line, similar to the work of Soto-Valero et al.~\cite{soto2021longitudinal}.

\subsubsection{Results}

\begin{figure}[b]
\vspace*{-5ex}
  \centering
  \includegraphics[width=0.95\linewidth]{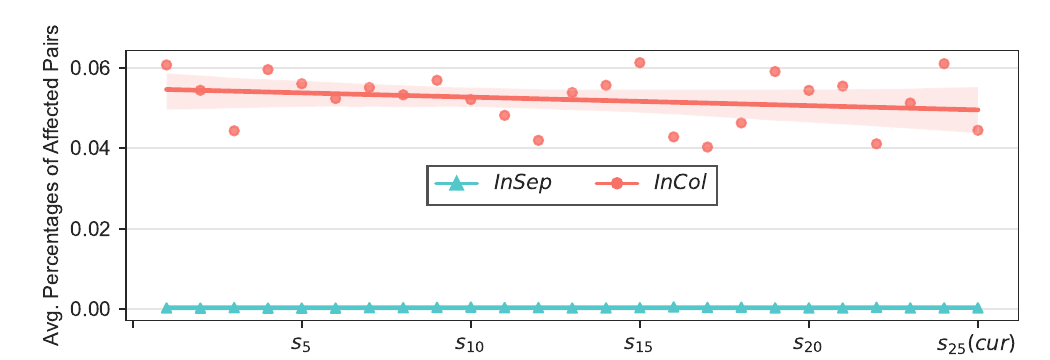}
  \caption{Stable evolutionary trends of InSep and InCol at pair level, averaged across all projects.}
  \label{fig: insep_incol_trend_pair}
\end{figure}

Fig.~\ref{fig: insep_incol_trend_pair} shows the evolution trend of $InSep$ and $InCol$ at pair level across all analyzed snapshots. 
Each data point represents an average percentage measured for each snapshot. The lines are linear regression functions, fitted to show the trend of $\mathit{InSep}$ and $\mathit{InCol}$ at a 95\% confidence interval.
From Fig.~\ref{fig: insep_incol_trend_pair}, the average percentages of $\mathit{InSep}$ remain stable across time.
For example, the percentage of $\mathit{InSep}$ in snapshot $s_{1}$ is 0.03\%, and by snapshot $s_{25}$ this value is still near to 0.03\%. 
For $\mathit{InCol}$, although we observe a slight decreasing tendency as systems grow, we find that such a tendency is not statistically significant (with $slop$ near to 0 and $p=0.26$). Thus, we conclude that overall, the percentages of smelly pairs for both forms remain stable over time, indicating that developers did not effectively intervene in \emph{PairSmell} issues within the analyzed time span.

\begin{figure}
  \centering
  \includegraphics[width=0.95\linewidth]{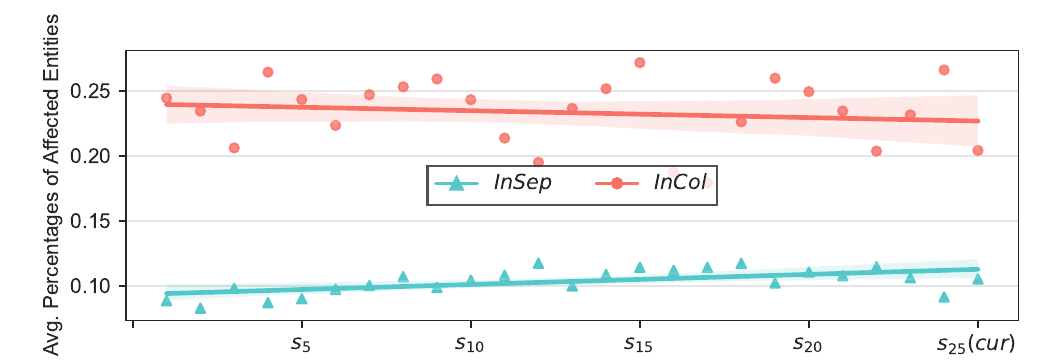}
  \caption{Evolutionary trends at entity level: increasing for InSep and stable for InCol, averaged across all projects}
  \label{fig: insep_incol_trend_entity}
\end{figure}

Fig.~\ref{fig: insep_incol_trend_entity} shows the evolution trend of entities involved in $\mathit{InSep}$ and $\mathit{InCol}$. Interestingly, we observe a clear and significant increasing tendency from the percentages of entities affected by $\mathit{InSep}$, despite the stable trend at pair level. 
Specifically, the proportion of entities affected by $\mathit{InSep}$ is 8.87\% in snapshot $s_{1}$ and 10.54\% in snapshot $s_{25}$ (increase $=$ 1.19x). This indicates that the number of entities newly affected by $\mathit{InSep}$ is generally higher than that of entities previously affected but no longer smelly, as systems grow.
On the other hand, a slight decreasing tendency can be observed from the percentages of entities with $\mathit{InCol}$.
Despite this, statistics show that such a tendency is not significant ($p=0.47$). 
We notice that the percentage of entities affected by $\mathit{InCol}$ is more variable ($SD=0.03$) and represents a larger share in comparison with that affected by $\mathit{InSep}$ ($SD=0.01$).

\begin{tcolorbox}[boxsep=0pt,left=2pt,right=2pt,top=3pt,bottom=2pt,arc = 0.1mm, boxrule = 0.3mm] \itshape
\textbf{RQ3 Summary:} 
The percentages of both $\mathit{InSep}$ and $\mathit{InCol}$ pairs do not diminish across the analyzed time, suggesting that PairSmell increases with system growth. Moreover, the percentages of entities affected by $\mathit{InSep}$ grow more noticeably, indicating a widespread and concerning phenomenon. 
\end{tcolorbox}

\section{Discussion and Implications}

Based on our empirical findings, this section discusses the discovery, management and further study of \emph{PairSmell}.
% we present implications and future work for practitioners and future researchers.

\subsection{Discovery of PairSmell}

\begin{table*}
\caption{Overview, characteristics, and processes of inductive and deductive approaches to discovering novel smells}
\label{tab: discovery}

\scriptsize
\begin{tabular}{p{20pt}p{220pt}p{220pt}}
\toprule
& \textbf{Inductive Approach} & \textbf{Deductive Approach}\\
\midrule
Overview & Smells are generalized from recurring observations in practice. & Smells are inferred from established premises.\\
Perspective & Now and past (problems observed in existing artifacts) & Now and future (possible problems based on theoretical premises)\\
Initiators & Practitioners, or researchers collaborating with practitioners & Researchers \\
\midrule
\multicolumn{1}{l}{\makecell[l]{Definition\\
Process}}& \makecell[l]{\tabitem  Observe and gather instances where the problem manifests. \\
\eg Configuration smells in~\cite{jafari2021dependency} are discovered based on vulnerable packages. \\
\tabitem  Identify the recurring characteristics across different instances.\\
\eg Authors~\cite{rahman2021security} observed recurring coding patterns as security smells.\\
\tabitem  Formulate a rule encapsulating the characteristics and justify its impacts. \\
\eg The impact of \textit{flaky test} is elucidated using real-world cases~\cite{yang2024lost}.\\
}
 & \makecell[l]{\tabitem  Formulate a theoretical premise that logically suggests specific problems.\\
 \eg Our premise is that decisions violating the apt ones could be problematic.\\
\tabitem  Describe what the problem looks like (\eg analysis units, problematic structure). \\
\eg PairSmell focuses on MRs and their corresponding deviations (Section~\ref{sec: smell definition}).\\
\tabitem  Justify the problem as a smell by highlighting its negative impacts on quality.\\
\eg Section~\ref{sec: smell identification} illustrates how PairSmell could impair a healthy modular structure. \\
}\\
\midrule
\makecell[lb]{Detection \\Method\\ Design} &  \makecell[l]{\tabitem  Define detection criteria (targets, indicators) based on inductive insights. \\
\eg \textit{Manual execution} is a configuration smell, except in deploy stages~\cite{vassallo2020configuration}. \\
\tabitem  Develop logical mechanisms (\eg algorithms) based on inductive data.\\
\eg Authors~\cite{vassallo2020configuration} set the \textit{Retry Failure} threshold based on known causes.\\
\tabitem  Implement and evaluate the detection tool with validation datasets\\ \eg Known or labeled smells~\cite{chen2023smelly,vassallo2020configuration} can serve as validation dataset.
} &  \makecell[l]{\tabitem  Translate the smell definition into quantifiable metrics based on the premise.\\ 
\eg PairSmell considers the MR between a pair as a key metric (Section~\ref{sec: smell identification}).\\
\tabitem  Develop logical mechanisms with theoretical consistency.\\
\eg Apt MRs, actual MRs, and their discrepancies help identify PairSmells. \\
\tabitem  Implement and evaluate the tool using validation datasets or manual review\\
\eg Premise ensures the detection results align with expectations (Section~\ref{sec: tool implementation}).
} \\
\midrule
\makecell[lb]{Usefulness\\
Assessment} & \multicolumn{2}{l}{\makecell[l]{\textbf{Prevalence:} Investigating its occurrences in practice to show its prevalence and importance. \\ \tabitem Detect the smell (using the developed tool) in real software projects and observe its frequencies and percentages.\\ Observations can guide project selection; \eg Jafari et al.~\cite{jafari2021dependency} excluded projects without ``package.json'' as it hinders pinpointing dependency smells. \\Premises help framing interpretation; based on our premise, Section~\ref{sec: prevalence} presents the number of apt MRs, actual MRs, and detected deviations (smells). }}
   \\
   & \multicolumn{2}{l}{\makecell[l]{\textbf{Consequences:} Provide empirical evidence demonstrating its impact on software quality. \\
\tabitem Collect quantitative data to show how the smell affects key quality metrics (\eg change-proneness), by comparing code artifacts with and without smells~\cite{abidi2021multi}.\\
Impacts can be hypothesized based on observations or premises. Our hypothesis (Section~\ref{sec: impacts}) stemmed from the unhealthy structure of \emph{PairSmells} vs. other pairs.\\
\tabitem Gather qualitative feedback from developers on the smell's impact on their workflow and codebase, \eg via issue reporting~\cite{vassallo2020configuration,rahman2021security} and surveys~\cite{jafari2021dependency,nardone2023video}. \\
Both inductive insights~\cite{nardone2023video} and established premises can inform the questions posed to developers during the evaluation.}}\\
\midrule
Benefits & 1) Enhanced practitioner acceptance; 2) Easy verification.  & 1) Broadened scope of smell knowledge; 2) Hastened smell discovery.  \\
Challenges & Delayed problem discovery could lead to higher maintenance costs. & Practitioners need to invest time in understanding the smell beforehand. \\ 
\bottomrule
\end{tabular}
\vspace*{-2.0ex}
\end{table*}

The discovery of new software smells, since Fowler's seminal work~\cite{fowler2018refactoring} on code smells, generally follows inductive and deductive approaches, as summarized in Table~\ref{tab: discovery}.
In inductive approach, recurring observations lead to the generalization of new smells~\cite{yang2024lost,jafari2021dependency,garcia2009identifying,Mo2015}, while in deductive approach, new smells are derived from theoretical premises~\cite{chen2023smelly,wong2011detecting}.
These approaches differ in their characteristics and processes of smell discovery, particularly in definition, detection, and assessment.
% , each offering unique benefits and addressing distinct challenges.
By reflecting on our research process and integrating methodologies reported in previous smell discovery studies (\eg~\cite{yang2024lost,lambiase2020just,vassallo2020configuration}), we derived Table~\ref{tab: discovery}. The characteristics and processes outlined in this table serve provide a reference and guide for future researchers and practitioners in proposing new smells.

\emph{PairSmell}, proposed in this study, is based on the premise that decisions violating appropriate or ideal ones could be problematic. By focusing on the MR perspective, it offers a granular yet fundamental aspect for inspecting modularity principles. Unlike the inductive approach, this deductive method (1) broadens the scope of smell knowledge by identifying potential issues previously unrecognized within the community, and (2) accelerates the discovery of new smells by proactively uncovering problems.

\subsection{Management of PairSmell}

\begin{figure}[b]
  \centering
  \includegraphics[width=0.92\linewidth]{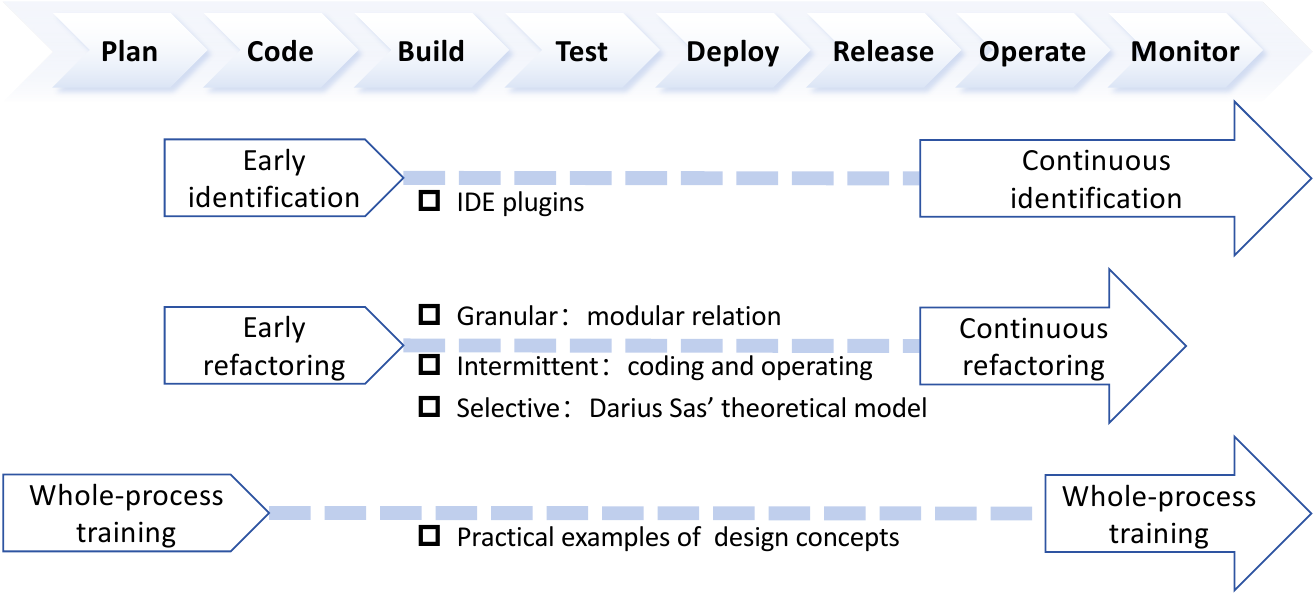}
  \caption{Management of PairSmell in DevOps development.}
  \label{fig: lifecycle}
\end{figure}

Our findings indicate that \emph{PairSmell} is significant for inspecting software modular structure (RQ1 and RQ2) and remains inadequately addressed (RQ3). This section discusses its management from three critical aspects (Fig.~\ref{fig: lifecycle}): identification,
resolution, and training (prevention). 

% . In the software development lifecycle, \eg DevOps~\cite{bass2015devops}, we envision better \emph{PairSmell} management through its identification, resolution, and prevention, as illustrated in Fig.~\ref{fig: lifecycle}.

\textbf{1) Early and continuous identification.}
A paramount benefit of \emph{PairSmell} is its ability to be detected automatically and pinpoint specific modular issues at the pair level (Section~\ref{sec: Pairwise Modular Smell}).
This capability, requiring minimal developer effort~\cite{kalhor2024systematic}, can be effectively integrated into IDE plugins for continuous assessment (coding, operating, and monitoring stages), similar to code smell tools such as SAT~\cite{romano2022static} and DARTS~\cite{lambiase2020just}. 
We envision that early and continuous identification of \emph{PairSmell} will improve the modular structure by enabling developers to uncover and address suboptimal modular decisions promptly.

\textbf{2) Granular, intermittent, and selective refactoring.}
Compared to prior smells, \emph{PairSmell} provides a granular but fundamental perspective for inspecting software modules.
To better manage it, we suggest the following refactoring strategies: 
Regarding \emph{how to refactor}, for an $\mathit{InSep}$ pair, developers could identify a single module to house both entities, \eg simply merging the entities into the module with the strongest connections to them.
For an $\mathit{InCol}$ pair, developers should examine the interactions between the two entities with other parts of the module, potentially separating the current module to establish clear boundaries.
Considering the main issue involves two entities, the corresponding refactoring operations, such as a single move class operation, are more actionable than those for coarse-grained smells. 
Regarding \emph{the timing of refactor}, engaging in \emph{intermittent floss refactoring}~\cite{noei2023empirical}—consistently integrating refactoring activities throughout the development process, with particular focus on the coding and operating stages—is recommended due to the relatively low resolution costs. In addition, it is advisable to address \emph{PairSmell} particularly during the \emph{vibrant} and \emph{growing} phases of projects to prevent detrimental co-changes in later development stages (Section~\ref{sec: impacts}).
% Developers should also prioritize refactoring to effectively manage \emph{PairSmells} in a project.
Regarding \emph{which smell to refactor first}, we suggest developers balance the severity of each smell instance---considering dependencies among the pair---and the remediation effort, such as the involved lines of code, following Darius Sas' theoretical model~\cite{sas2023architectural}.

\textbf{3) Whole-process modular training.}
The widespread occurrence of \emph{PairSmells} in numerous projects underscores the need for improved training in software design.
We acknowledge that teaching design concepts is a challenge~\cite{CAI2023107322}; after all, many design principles such as SOLID~\cite{martin2018clean}, DRY~\cite{thomas2019pragmatic}, and SoC~\cite{laplante2022every} can often seem too abstract.
However, educators can demystify these concepts with practical examples of modular smells, such as \emph{PairSmells}, as demonstrated in Section~\ref{sec: Pairwise Modular Smell}.
Developers can also enhance their understanding of modular design by actively identifying \emph{PairSmells} in their projects and conducting detailed analyses to mitigate these issues.

\subsection{Further Study of PairSmell}

\noindent
\textbf{Empirical evidence suggests that \emph{PairSmell} could undermine the ideal modular structure of a project during software maintenance. Future studies should continue to gather feedback from practitioners on \emph{PairSmell}.}

As we find in RQ2, $\mathit{InSep}$ correlates with increased co-changes across modules; while $\mathit{InCol}$ pairs exhibit fewer, suggesting reduced module coherence.
Both observations violate the modular design principles~\cite{arvanitou2015introducing,martin2018clean} and undermine software maintenance and change.
Our findings provide preliminary insights into the usefulness of \emph{PairSmell} for inspecting software modular quality.
Nevertheless, future studies should further evaluate \emph{PairSmell} by examining its relevance to developers, to better understand its impact from the developers' perspective.
One possible method is collecting developers feedback by opening \emph{PairSmell} issues in issue trackers, similar to Vassallo et al.'s approach~\cite{vassallo2020configuration}. The validity of \emph{PairSmell} could be confirmed if developers not only agree with the issues but also take actions to address them.

\section{Threats to Validity}

\emph{Internal validity} could be threaten by factors that influence smell identification. 
A relevant threat is that the inferred apt MRs might be biased by individual tools.
Research on consensus clustering find that low-quality base clusterings can degrade the quality of final ensemble solutions~\cite{golalipour2021clustering}. The modularization tools we selected might yield poorly structured solutions and potentially unreasonable MR design. To minimize this threat, we follow a set of rigorous criteria to select tools that have demonstrated promising results. 
We avoid non-deterministic tools (\eg Bunch~\cite{mitchell2006automatic}) which could introduce their own chance factors.
Additionally, we choose 4 distinct tools to further reduce the likelihood of biases by individual tools while maintaining an acceptable overhead. 
On the other hand, we use a system's lowest level folders as its existing modules for extracting actual MRs. While folders can reflect the development architectural view, not all folders are meaningful from architecture's perspective~\cite{zhang2023software}. For example, many C/C++ projects organize header files and cpp files into separate folders, and smells suggesting to move them into the same folder can be examples of \emph{false positives}.
% that offers no architectural insights. 
% Future studies should explore how this module extraction method affects \emph{PairSmell} detection, such as \emph{false positive}.

\emph{External validity} could be threatened, impairing the generalization of our findings. We are aware that our results may not be generalizable to other projects since all 20 studied projects are open source. To minimize this threat, a set of criteria are used to select projects varying widely across different domains and project characteristics. Future studies are encouraged to replicate our research on other projects in different settings.

\emph{Construct validity} could be threatened by possible imprecision in our measurements. This can be related to possible mistakes in our tool's implementation, beyond what we could discover by testing. We performed extensive manual examination to mitigate this threat. In addition, the dependent variables used to measure co-changes, \ie COR, CCO, and DOR, are defined based on the evidence commonly used for studying co-change relations~\cite{mo2023exploring,jin2020exploring} and software maintenance~\cite{mo2016decoupling}, 
and thus can be considered constructively valid.

\section{Related Work}

\noindent
\textbf{Software Modularization Techniques.} Over the past two decades, numerous techniques have been developed to restructure a large software system into smaller, and more manageable subsystems~\cite{sarhan2020software}.
These techniques typically conceptualize modularization as an optimization problem, seeking an optimal solution to refactor the original modules.
% based on predefined optimization objectives. 
The most commonly used optimization objectives are intraconnectivity (\emph{high cohesion}) and interconnectivity (\emph{low coupling}), \eg~in \cite{teymourian2020fast,yang2022enhancing,pourasghar2021graph,mitchell2006automatic}.
Additionally, some researchers incorporate refactoring effort, such as the number of changes~\cite{schroder2021search}, as an objective to minimize the effort required for modularization.
However, an industrial case study~\cite{schroder2021search} reveals that completely modularizing an entire system remains prohibitively expensive and thus impractical, given the extensive size of the code base.
\textit{Instead of seeking to restructure an entire system, this paper aims to integrate the intelligence of multiple modularization techniques to deduce promisingly appropriate MR designs and identify opportunities that necessitate refactoring.}

A few modularization techniques also focus on MR decisions. Erdemir and Buzluca~\cite{erdemir2014learning} calculated the probability of two entities being within the same module and utilized this information to promote subsequent modularization. Chong and Lee~\cite{chong2017automatic} obtained two constraints---an entity pair \textit{must be} and \textit{must not be} within a module---as the foundation for constraint-based clustering to enhance modularization solutions. \textit{Our study diverges from these research not only in objectives but also in the methodologies used to determine apt MRs.}

\noindent
\textbf{Metrics and Smells for Identifying Modularity Issues.} Identifying and alarming modularity-related `issues' is an essential objective for many architecture analysis activities, such as architectural quality measurement~\cite{al2012precise} and architectural smell detection~\cite{mo2019architecture,liu2024prevalence}. 
Architecture metrics, including modularity and maintainability measures~\cite{mo2016decoupling}, aim to assess the extent
to which a software system is maintainable. 
In addition, numerous metrics of coupling~\cite{almugrin2016using} and cohesion~\cite{athanasopoulos2014cohesion} can be employed to identify quality issues at the module level, for example, MCI~\cite{zhong2023measuring} for microservice coupling.

Architectural smells represent structural problems that negatively influence software evolution~\cite{xiao2021detecting,mo2019architecture} to indicate refactoring opportunities in subsequent development. Since Joshua Garcia's definition~\cite{garcia2009identifying}, numerous types of architectural smells have been introduced within the community.
To the best of our knowledge, smells relevant to \emph{PairSmell} include: \emph{Modularity Violation}~\cite{wong2011detecting}, referring to two components that consistently change together but belong to separate modules; 
\emph{Implicit Cross-module Dependency}~\cite{Mo2015}, indicating two structurally independent modules that frequently change together in the revision history;
\emph{Co-change Coupling}~\cite{le2018empirical}, where changes to one component require changes in another component.
\textit{Compared to these smells, the novelty of PairSmell manifests in two aspects: (1) PairSmell is defined at the fine-grained pair level, thus providing more actionable insights to enhance existing software modules than those targeting the module or component level; (2) while the above smells focus on the deviation between modular structure and historical revisions, PairSmell concerns deviation in the modular structure from the apt or ideal design decisions, offering a broader perspective than the existing smells.
}

\section{Conclusion}

Focusing on the granular pair-level, we introduce a novel architectural smell that reveals modular issues due to deviations from consensus modular decisions.
With the empirical study on 20 open source projects, we explore the prevalence and consequences of such smells. Our study presents solid evidence that the impact of such smells is nontrivial, but unordinarily high in practice by comparing the pairs with and without smells.

Our study benefits software research and practice by: (1) introducing a novel type of smell for inspecting software modular structure, (2) providing empirical evidence of its prevalence and consequences, and (3) suggesting how software modular activities can be enhanced---augmented with \emph{PairSmell}'s identification, resolution, and training.
This smell envisions contributing to software engineering by enabling more targeted and effective module enhancements.

% \section*{Acknowledgment}

% The preferred spelling of the word ``acknowledgment'' in America is without 
% an ``e'' after the ``g''. Avoid the stilted expression ``one of us (R. B. 
% G.) thanks $\ldots$''. Instead, try ``R. B. G. thanks$\ldots$''. Put sponsor 
% acknowledgments in the unnumbered footnote on the first page.

\bibliographystyle{ieeetr}
\bibliography{references}

\end{document}